\documentclass[twocolumn]{aastex701}
\usepackage{amsmath}
\defcitealias{2026A&A...706A..79P}{Paper~I}
\defcitealias{2026A&A...706A.357P}{Paper~II}

\begin{document}

\title{Reframing the Wide Eccentric Binary Problem: \\ Eccentricity as a Probe of Mass-transfer Physics}

\author[orcid=0009-0005-3096-1114,gname=Adam, sname= Parkosidis]{A. Parkosidis}
\affiliation{Anton Pannekoek Institute for Astronomy, University of Amsterdam, Amsterdam 1098 XH, The Netherlands}
\email[show]{a.parkosidis@uva.nl}  

\author[orcid=0000-0000-0000-0002,gname=Silvia, sname=Toonen]{S. Toonen} 
\affiliation{Anton Pannekoek Institute for Astronomy, University of Amsterdam, Amsterdam 1098 XH, The Netherlands}
\email[]{toonen@uva.nl} 

\author[orcid=0000-0003-1009-5691,gname=Eva,sname=Laplace]{E. Laplace}
\affiliation{Institute of Astronomy, KU Leuven, Celestijnenlaan 200D, B-3001 Leuven, Belgium}
\affiliation{Leuven Gravity Institute, KU Leuven, Celestijnenlaan 200D, box 2415, 3001 Leuven, Belgium}
\affiliation{Anton Pannekoek Institute for Astronomy, University of Amsterdam, Amsterdam 1098 XH, The Netherlands}
\email[]{eva.laplace@kuleuven.be}

\author[orcid=0000-0001-6339-6768, gname=Veronika,sname=Schaffenroth]{V. Schaffenroth}
\affiliation{Thüringer Landessternwarte Tautenburg, Sternwarte 5, 07778 Tautenburg, Germany}
\email[]{schaffenroth@tls-tautenburg.de}

\begin{abstract}

Observations of wide post-interaction binaries show an unexpected feature; orbital eccentricity, which is not understood theoretically. A promising resolution to this long-standing puzzle is eccentric mass transfer (MT). Here the first complete framework for MT in orbits with arbitrary eccentricity, the general mass-transfer (GeMT) model, is confronted  with the latest observations of hot subdwarfs of spectral type B (sdB) with main-sequence (MS) companions in wide orbits. SdBs are excellent benchmarks for binary evolution models, since their progenitors provide unique constraints on their formation histories. We isolate the effects of eccentric MT from other physical process and we show that it explains the observed orbital-parameter distributions and their correlations seen in wide sdB+MS binaries. To quantitatively constrain their orbital parameters, stellar
evolution and tides should be included in future studies, yet it is encouraging that the GeMT model provides the first mechanism that qualitatively reproduces the observed distributions. We further demonstrate that post-MT eccentricities depend directly on key MT parameters, including transferred mass, accretion efficiency, and angular momentum loss. Given the multitude of eccentric post-MT binaries with components ranging from low- to high-mass stars to compact objects, we propose that post-MT eccentricities offer a new window onto binary evolution, presenting a powerful tool to constrain highly uncertain binary-evolution parameters and mass-transfer formation histories across diverse populations. Post-MT eccentricity should therefore be embraced as a key observable, rather than treated as a problem to be corrected.

\end{abstract}



\section{Introduction}\label{sec:one}

Classical binary theory predicts that tidal interaction should circularizes orbits before the onset of mass transfert (MT) \citep{2002MNRAS.329..897H,1996A&A...309..179P}. Observations, however, reveal that nonzero eccentricities are common in wide post-interaction systems \citep[e.g.,][]{2018A&A...620A..85O,2019A&A...626A.127J,2019A&A...626A.128E,2019MNRAS.482.4592V,2021A&A...653A.120D,2022A&A...658A.122M,2024MNRAS.52711719Y,2024ApJ...962...70K,2025A&A...701A...9M} as well as in systems undergoing MT \citep{1999AJ....117..587P,2005A&AT...24..151R}. Furthermore, the range of observed eccentricities increases with orbital period \citep[e.g., Figure~8 in][]{2024MNRAS.529.3729S}. This pattern is evident across several post-interaction binaries, and is not confined to a specific population. This discrepancy between classical binary theory and observations is known as the eccentricity problem. 

Several mechanisms have been proposed, but synthetic models still fail to match the global orbital properties of post-interaction binaries: (a) Tidally enhanced wind mass loss \citep{2000A&A...357..557S,2008A&A...480..797B} leads to eccentric helium-white-dwarf binaries but not hot subdwarfs of spectral type B (sdB) systems, because extreme mass loss prevents helium ignition \citep{2015A&A...579A..49V}. (b) Circumbinary-disk (CBD) interactions can pump the eccentricity to about $0.2$, and thus cannot explain more eccentric post-asymptotic giant branch (post-AGB) binaries \citep{2020A&A...642A.234O}. Importantly, such interactions produce higher eccentricities at shorter orbital periods, contradicting observations \citep{2013A&A...551A..50D,2015A&A...579A..49V,2018MNRAS.474..433D}. (c) White dwarf kicks \citep{2010A&A...523A..10I,2025arXiv250821805C} may increase eccentricity, but are irrelevant for sdB binaries, and the kick mechanism also remains unclear. (d) Mergers in triples \citep{2009ApJ...697.1048P} or dynamical interactions with a tertiary companion \citep{2020A&A...640A..16T} can produce eccentric binaries. In the first scenario, the surviving binary consists of the merger product and the original tertiary companion. In the second scenario, the Lidov-Kozai mechanism drives eccentricity growth in the inner binary. Nevertheless, it is unlikely that triple interactions alone can account for all the eccentric orbits observed across the entire population of post-interaction binaries. In summary, none of the above mechanisms can (1) easily reproduce the orbital parameters of the observed eccentric post-interaction binaries and (2) naturally explains the observed increase of maximum eccentricity with orbital period (Figure~\ref{fig:period_eccentricity_plane}), a trend that appears to hold across the entire post-interaction binary population.

In this letter, we utilize the general mass-transfer framework \citep[GeMT;][]{2026A&A...706A..79P,2026A&A...706A.357P} to investigate how eccentric MT shapes the orbits of post-mass-transfer binaries.By isolating eccentric MT from other processes, we show that GeMT model explains the observed orbital-parameter distributions and their correlations seen in wide sdB+MS binaries. More broadly, it predicts that post-MT eccentricities are largely independent of orbital period but depend directly on key MT parameters such as the amount of transferred mass, accretion efficiency, and angular-momentum loss, thereby clarifying the broader role of eccentricity in MT evolution.

\section{Wide sdB binaries: benchmark for binary evolution}\label{sec:two}

SdBs are compact, high-temperature stars with  $\mathrm{T}_{\rm eff} \gtrsim 20000$ K, understood as core He-burning objects with extremely thin hydrogen envelopes \citep{2009ARA&A..47..211H,2016PASP..128h2001H,2025ARA&A..63..467M}. Most observed sdBs have companions and the current consensus is that they form in binaries. SdB binaries with low-mass M-type or brown dwarfs (BDs) companions are found in short periods of $0.05 \lesssim P_{\rm orb} \lesssim 1.6\, \mathrm{days}$ \citep{2019A&A...630A..80S,2022A&A...666A.182S}, and likely form via unstable MT \citep{2002MNRAS.336..449H,2003MNRAS.341..669H} followed by a common-envelope (CE) phase \citep{1976IAUS...73...75P}. In contrast, sdB binaries with FGK-type main-sequence (MS) companions are observed at long-period $500 \lesssim P_{\rm orb} \lesssim 1400 \, \mathrm{days}$ \citep{2018MNRAS.473..693V}, and
thought to form via stable MT \citep{2002MNRAS.336..449H,2003MNRAS.341..669H,2013MNRAS.434..186C}. Finally, sdBs with white dwarf companions have periods from of 0.5 hours to $\sim 30$ days and are thought to have undergone an initial phase of stable MT, followed by an unstable MT episode once the sdB companion evolves into a red giant \citep{2009ARA&A..47..211H}.

In the stable MT formation channel, the sdB progenitor overflows its Roche lobe during red giant evolution, losing most of its hydrogen-rich envelope but still igniting helium during or shortly after envelope ejection. Meanwhile, the companion accretes mass and angular momentum and spins up \citep{1991A&A...241..419P}. Low-mass progenitors (i.e.  $M \lesssim 1.5 \, \mathrm{M}_{\odot}$) yield sdB masses close to the typical degenerate core-helium-flash mass of $\sim 0.47 \, \mathrm{M}_{\odot}$, while a broader range  $0.35 - 0.65 \, \mathrm{M}_{\odot}$ is possible for more massive progenitors \citep{2002MNRAS.336..449H,2003MNRAS.341..669H,2024MNRAS.52711184A}.

In a binary population synthesis study, \cite{2003MNRAS.341..669H} predicted that stable MT produces wide sdBs+MS systems with orbital periods up to $\sim 500\, \mathrm{days}$. Classical tidal theory further predicts that binaries with evolved giant donors should be circular by the end of RLOF \citep[e.g.,][]{1996A&A...309..179P,2002MNRAS.329..897H}. Observations, however, contradict both expectations: the observed orbital periods are considerably longer \citep[see][]{2013MNRAS.434..186C}, and nearly all systems are eccentric. \citet{2013MNRAS.434..186C} reproduced the longer periods--up to $P_{\rm orb} \sim 1600\, \mathrm{days}$--by including atmospheric RLOF and treating angular momentum loss (AML) in greater detail, but the eccentricity remains puzzling.

A well-constrained sample of 24 wide sdB+MS binaries is presented in Table~\ref{tab:observables}.\footnote{Including PG1701+359 \citep{2013ApJ...771...23B} gives 25 systems, but the mass ratio $q_{\rm obs}$ of the system is unknown.} This sample reveals structure in their demographics: 
\begin{enumerate}
    \item Longer $P_{\rm orb}$ at lower mass ratios, $q_{\rm obs} = M_{\rm sdB}/ M_{\rm MS}$ \citep{2020A&A...641A.163V},
    \item Higher eccentricities, $e$, at lower $q_{\rm obs}$ \citep{2022A&A...658A.122M},
    \item Higher $e$ at longer $P_{\rm orb}$ \citep{2017A&A...605A.109V},
\end{enumerate}
as shown in Figures~\ref{fig:streamplot_dPdq}, \ref{fig:streamplot_dedq}, and~\ref{fig:streamplot_dedP}, respectively. Distributions (1) and (3) appear bimodal \citep{2020A&A...641A.163V}, with a ``main branch'' (blue circles; $P_{\rm orb} \gtrsim 750\, \mathrm{days}$) and a ``secondary branch'' (green circles; $P_{\rm orb} \lesssim 750\, \mathrm{days}$) following the same trend at different $P_{\rm orb}$ (Figures~\ref{fig:streamplot_dPdq} and \ref{fig:streamplot_dedq}). Distribution (2) is unimodal (Figure~\ref{fig:streamplot_dedP}).

\cite{2020A&A...641A.163V} attribute correlation (1) for the main branch to Galactic metallicity evolution combined with MT. The origin of the secondary branch remains unexplained. Correlations (2) and (3) are even more puzzling because eccentricity is not expected. \cite{2015A&A...579A..49V} invoked phase-dependent RLOF plus a circumbinary disk to generate eccentricities, but this mechanism predicts higher $e$ at shorter $P_{\rm orb}$ \citep[see also][]{2013A&A...551A..50D,2018MNRAS.474..433D,2020A&A...642A.234O}, contrary to correlation (3). No theoretical explanation currently accounts for correlation (2) or its unimodal character. In summary, current theory cannot easily explain and simultaneously reproduce the observed trends of wide sdB+MS systems.

We use the GeMT framework to predict the orbital evolution of mass-transfering binaries that form wide sdBs+MS binaries. Speficially, we consider an eccentric binary with donor mass $M_{\rm don}$ , accretor mass $M_{\rm acc}$, and mass ratio $q=M_{\rm don}/M_{\rm acc}$. The orbit has a semimajor axis $a$, an eccentricity $e$, and a period $P_{\rm orb}$. Assuming that the angular momentum stored in the stellar spins is negligible compared to the orbital angular momentum (i.e., limit of point masses), the secular orbital evolution is given by
\begin{eqnarray}
 &\frac{\langle \dot{a} \rangle}{a} = -\frac{2 \langle \dot{M}_{\rm don} \rangle}{M_{\rm don}} \frac{f_{a}(e,x)}{f_{\dot{M}_{\rm don}}(e,x)}  \Biggl(1-\beta q-(1-\beta)\frac{(\gamma+\frac{1}{2})q}{1+q}\Biggr), \label{eq:orbit_averaged_semimajor_axis} \\ 
 &\langle \dot{e} \rangle = -\frac{2 \langle \dot{M}_{\rm don} \rangle}{M_{\rm don}} \frac{f_{e}(e,x)}{f_{\dot{M}_{\rm don}}(e,x)} \Biggl(1-\beta q-(1-\beta)\frac{(\gamma+\frac{1}{2})q}{1+q}\Biggr),\label{eq:orbit_averaged_eccentricity} \\
 &\langle \dot{\omega} \rangle = 0, \label{eq:orbit_averaged_argument_of_periapsis} \nonumber
\end{eqnarray}
where $\beta$ is the fraction of the transferred mass that is accreted,\footnote{Traditionally, $\beta$ denotes the fraction of mass transferred that is ejected near the accretor (e.g., \citet{1997A&A...327..620S}). Here, we follow the notation introduced in Section 7.2, pages 9–12, of \href{https://www.astro.ru.nl/~onnop/education/binaries_utrecht_notes/Binaries_ch6-8.pdf}{lecture notes on binary star evolution} by Onno Pols.} $\gamma$ measures AML efficiency in nonconservative MT \citep{2026A&A...706A.357P}, $x \equiv R_{\rm L}^c/R_{\rm don}$ represents the level at which the physical donor radius, $R_{\rm don}$, overflows the Roche-lobe equivalent radius, $R_{\rm L}^c$, for a circular orbit, and $f_{\dot{M}_{\rm don}}(e,x)$, $f_{a}(e,x)$, $f_{e}(e,x)$, are dimensionless functions given explicitly in Appendix E of \citep{2026A&A...706A..79P}.

We confront the GeMT-model predictions with the observed wide sdB+MS sample (Table~\ref{tab:observables}). These systems are chosen because (1) they are consistent with having undergone only a single phase of stable MT \citep[][]{2009ARA&A..47..211H,2016PASP..128h2001H,2024arXiv241011663H}, (2)  most sdB masses should cluster near the degenerate core-helium-flash mass of $0.47 \mathrm{M}_{\odot}$ \citep{2009ARA&A..47..211H,2016PASP..128h2001H}, and (3) reaching this mass requires that RLOF begins close to the tip of the red giant branch \citep[TRGB;][]{2002MNRAS.336..449H,2003MNRAS.341..669H}, tightly constraining the radius at MT onset. Consequently, wide sdB+MS systems impose strong constraints on their formation histories and provide a powerful test of MT physics.

\begingroup
\begin{figure*}[!htp]
  {\includegraphics[width=0.99\linewidth]{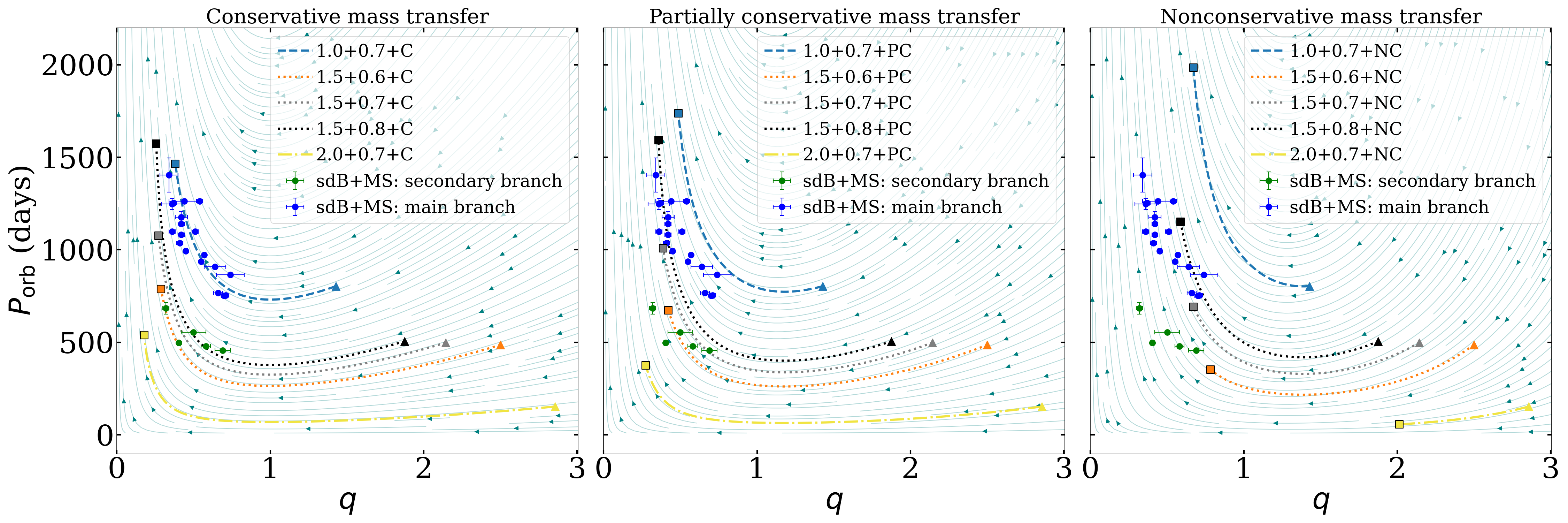}}%
  \caption{Evolution of the orbital period, $P_{\rm orb}$, as a function of mass ratio, $q = M_{\rm don}/M_{\rm acc}$.  Dashed, dotted and dash-dotted lines illustrate the evolution for $1~\mathrm{M}_{\odot}$, $1.5 \, \mathrm{M}_{\odot}$, and $2 \, \mathrm{M}_{\odot}$ donors, respectively. Triangles and squares indicate the initial (i.e., at the RLOF onset) and final (i.e., post-mass-transfer) positions of the models, respectively. Teal arrows correspond to Equation~\eqref{eq:dPdq} and indicate the evolution of $P_{\rm orb}$ as mass transfer proceeds (i.e., as $q$ decreases) for $e=0.001$ and $x=0.99$.  From left to right: Conservative mass transfer ($\beta = 1$), partially conservative mass transfer ($\beta = 0.5$) under isotropic reemission ($\gamma = q$), and nonconservative mass transfer ($\beta = 0$) under isotropic reemission ($\gamma = q$). Blue and green circles represent observed post-mass-transfer sdB+MS systems (Table~\ref{tab:observables}). Blue circles correspond to the main branch ($P_{\rm orb} \gtrsim 750\, \mathrm{days}$) and the green circles to the secondary branch ($P_{\rm orb} \lesssim 750\, \mathrm{days}$). The parameters of the models at the onset of mass transfer (triangles) are listed in Table~\ref{tab:integration_parameters}. The post-mass-transfer parameters of the models (squares) are listed in Table~\ref{tab:end_of_MT}.}\label{fig:streamplot_dPdq}
\end{figure*}
\endgroup

\begingroup
\begin{figure*}[!htp]
  {\includegraphics[width=0.99\linewidth]{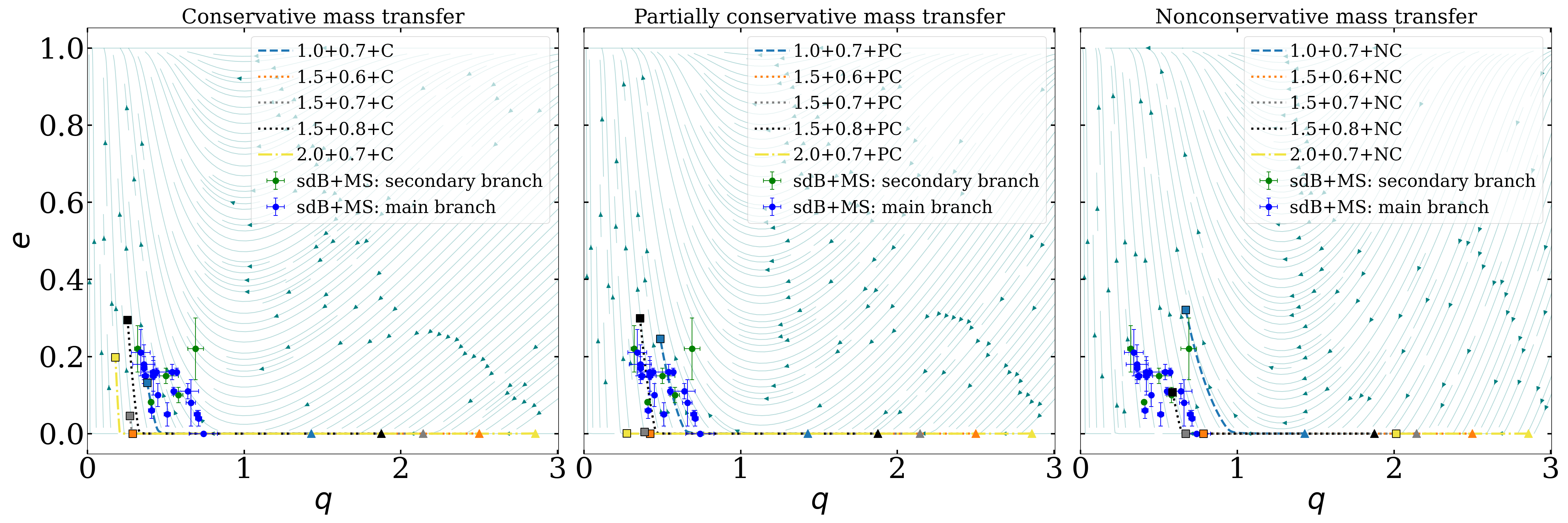}}%
  \caption{Similar to Figure~\ref{fig:streamplot_dPdq}, but the vertical axis now illustrates the orbital eccentricity, $e$, and the teal arrows correspond to Equation~\eqref{eq:dedq}.}
  \label{fig:streamplot_dedq}
\end{figure*}
\endgroup

\begin{figure}[!htbp]
    \centering
    \includegraphics[width=\linewidth]{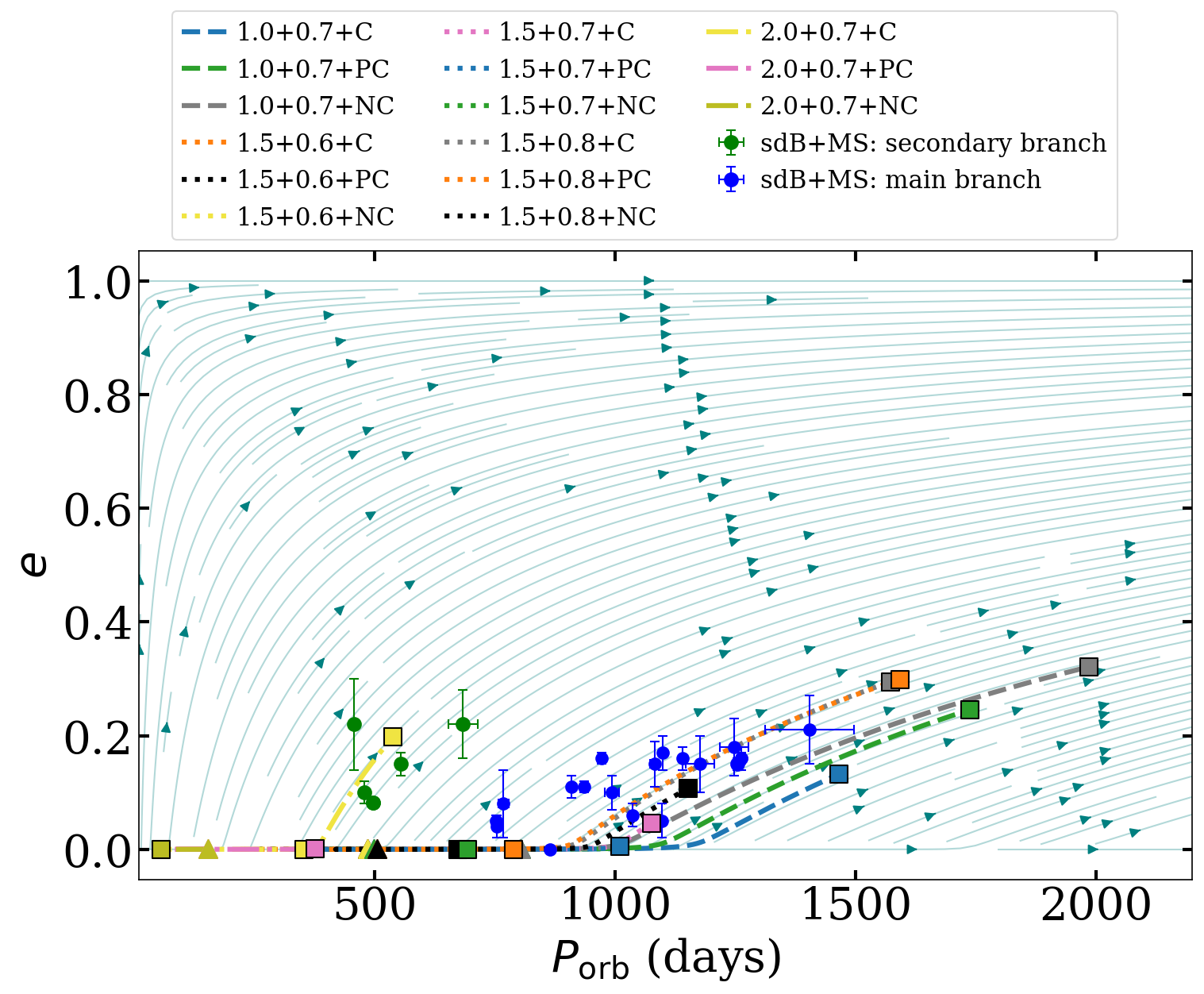}
    \caption{Similar to Figure~\ref{fig:streamplot_dedq}, but the horizantal axis now illustrates the orbital period, $P_{\rm orb}$ and the teal arrows correspond to Equation~\eqref{eq:dedP}.}
    \label{fig:streamplot_dedP}
\end{figure}

\section{Results}\label{sec:three}

\subsection{Comparison to observations}\label{subsec:comparison}

In this section, we isolate the effects of MT via eccentric RLOF from other physical processes, such as stellar evolution and tides, in order to highlight its impact on the orbital evolution. We numerically integrate Equations~\eqref{eq:orbit_averaged_semimajor_axis} and \eqref{eq:orbit_averaged_eccentricity} to compare the predicted orbital evolution of systems that form sdB+MS binaries with the observed ones listed in Table~\ref{tab:observables}. We select three typical sdB progenitors, $1 \, \mathrm{M}_{\odot}$, $1.5 \, \mathrm{M}_{\odot}$, and $2 \, \mathrm{M}_{\odot}$ that initiate RLOF at the tip of the RGB in nearly circular orbits ($e = 0.001$; see also Appendix~\ref{app:sensitivity}). Such seed eccentricities are expected for Roche-lobe-filling giants with convecting envelopes \citep{1992RSPTA.341...39P,2024MNRAS.534..455C}. The donor radii and helium cores near the TRGB are taken from the detailed MESA calculations of \citet{2023A&A...669A..45T}. 

In total, we construct fifteen models by varying the initial accretor mass and semimajor axis such that all systems initiate RLOF at periapsis, and by varying the accretion efficiency $\beta$. Specifically, we investigate three cases: (a) conservative MT ($\beta=1$), (b) partially conservative MT ($\beta=0.5$) under isotropic reemission ($\gamma = q$), in which the ejected mass carries away the specific angular momentum of the accretor, and (c) nonconservative MT ($\beta=0$) under isotropic reemission ($\gamma = q$). We label the models by the initial donor mass $M_{\rm don}$, initial accretor mass $M_{\rm acc}$, and whether MT is conservative (C), partially conservative (PC) or nonconservative (NC). All model parameters at the onset of RLOF are listed in Table~\ref{tab:integration_parameters}.

The orbital evolution of the models along the $q - P_{\rm orb}$, $q - e$, and $P_{\rm orb} - e$ planes is presented in Figures~\ref{fig:streamplot_dPdq} to \ref{fig:streamplot_dedP}, respectively. In all cases, MT proceeds until the envelope has been removed; for $M_{\rm don}=2\, \mathrm{M}_{\odot}$ until $M_{\rm don} = 0.41\,\mathrm{M}_\odot$, and in all other cases until $M_{\rm don} = 0.47\,\mathrm{M}_\odot$, with final orbital parameters listed in Table~\ref{tab:end_of_MT}. Model endpoints (squares in Figures~\ref{fig:streamplot_dPdq} to \ref{fig:streamplot_dedP}) are compared to observed wide sdB+MS systems (circles in Figures~\ref{fig:streamplot_dPdq} to \ref{fig:streamplot_dedP}) across all three planes, subject to physical constraints. 

\begin{deluxetable}{cccccc}
\tablecaption{ Orbital parameters at the end of RLOF.}
\label{tab:end_of_MT}
  \tablehead{ \colhead{Model} & \colhead{$M_{\rm don}$} & \colhead{$M_{\rm acc}$} & \colhead{$q$} & \colhead{$P_{\rm orb}$} & \colhead{$e$}\\
  & (M$_{\odot}$) & (M$_{\odot}$) & & (days) &  }
  \startdata
  1.0+0.7+C & 0.47 & 1.23 & 0.382 & 1464.95 &  0.132 \\
  1.0+0.7+PC & 0.47 & 0.965 & 0.487 & 1737.6 & 0.246\\
  1.0+0.7+NC & 0.47 & 0.7 & 0.671 & 1984.34 &  0.321  \\
  \hline
  1.5+0.8+C & 0.47 & 1.83 & 0.257 & 1573.15 &  0.295  \\
  1.5+0.7+C & 0.47 & 1.73 & 0.272 & 1075.94 &  0.046  \\
  1.5+0.6+C & 0.47 & 1.63 & 0.288 & 788.05 &  0.0001  \\
  1.5+0.8+PC & 0.47 & 1.315 & 0.357 & 1591.65 &  0.299  \\
  1.5+0.7+PC & 0.47 & 1.215 & 0.387 & 1009.33 &  0.005  \\
  1.5+0.6+PC & 0.47 & 1.115 & 0.422 & 673.44 &  0.00005  \\
  1.5+0.8+NC & 0.47 & 0.8 & 0.588 & 1152.39 &  0.107  \\
  1.5+0.7+NC & 0.47 & 0.7 & 0.671 & 692.81 & 0.00003 \\
  1.5+0.6+NC & 0.47 & 0.6 & 0.783 & 352.95 &  0.000007  \\
  \hline
  2.0+0.7+C & 0.41 & 2.29 & 0.179 & 538.57 &  0.198  \\  
  2.0+0.7+PC & 0.41 & 1.495 & 0.274 & 375.98 &  0.0004  \\
  2.0+0.7+NC & - & - & - & - &  -\\
  \hline
  \enddata
  \tablecomments{Models are labeled by the donor mass $M_{\rm don}$, the accretor mass $M_{\rm acc}$, and whether MT is conservative (C), partially conservative (PC) or nonconservative (NC). The orbital parameters correspond to the endpoints of the models and are shown as squares in Figures~\ref{fig:streamplot_dPdq} to \ref{fig:streamplot_dedP}. Model 2.0+0.7+NC terminates during MT because the system formally merges (i.e., $R_{\rm don} \geq a(1-e$)).}
\end{deluxetable}

Overall, after eccentric MT, the endpoints of the models are near the observed systems (see squares in Figures~\ref{fig:streamplot_dPdq} to \ref{fig:streamplot_dedP}). Systems with $1~\mathrm{M}_{\odot}$ progenitors initiating RLOF near $P_{\rm orb} \sim 800\, \mathrm{days}$ match the main branch best under conservative transfer. $1.5 \, \mathrm{M}_{\odot}$ progenitors starting RLOF at $P_{\rm orb} \in [486,505]\, \mathrm{days}$ can reproduce the main branch for both conservative and nonconservative cases, and the secondary branch when paired with low-mass companions ($q \gtrsim 2.5$). $2~\mathrm{M_\odot}$ progenitors reproduce secondary-branch periods and eccentricities only for conservative transfer, otherwise merging. Overall, main-branch systems favor $1–1.5~\mathrm{M_\odot}$ progenitors with moderate mass ratios; secondary branch requires heavier progenitors ($\gtrsim 1.5 \; \mathrm{M}_{\odot}$) and/or larger initial mass ratios (i.e., low-mass companions) to match observations. The observed eccentricites are reproduced by all three progenitors with a notable trend; for a given sdB progenitor mass and accretion efficiency $\beta$, a lower initial $q$ result in higher $e$ and post-MT $q$ (see models with $1.5 \, \mathrm{M}_{\odot}$ progenitor in Figures~\ref{fig:streamplot_dPdq} to \ref{fig:streamplot_dedP}).

\subsection{Interpretation} 

In Section~\ref{subsec:comparison}, we showed that the GeMT model predicts an orbital evolution that is qualitatively consistent with the observed orbital parameters of wide sdB+MS binaries. Here, we demonstrate that it also explains their observed orbital-parameter correlations (1)-(3) introduced in Section~\ref{sec:two} and shown in Figures~\ref{fig:streamplot_dPdq} to \ref{fig:streamplot_dedP}. 

To interpret these correlations, we first derive a set of intuitive equations. The GeMT framework predicts that $q$ is the dominant parameter determining orbital evolution during MT. Its rate of change follows
\begin{equation}\label{eq:q_dot}
    \frac{\dot{q}}{q} = (1+\beta q) \frac{\dot{M}_{\rm don}}{M_{\rm don}},
\end{equation}
and using Equations~21, 24, and 28 from \citetalias{2026A&A...706A..79P}, its orbit-averaged rate of change is
\begin{equation}\label{eq:q_dot_sec}
    \frac{\langle \dot{q} \rangle}{q(1+\beta q)} = \frac{\langle \dot{M}_{\rm don}\rangle}{M_{\rm don}},
\end{equation}
where $\langle ... \rangle$ denotes orbit-averaged quantities. Combining Equations~\eqref{eq:q_dot_sec}, \eqref{eq:orbit_averaged_semimajor_axis}, \eqref{eq:orbit_averaged_eccentricity}, and Kepler's third law,\footnote{When substituting Kepler's third law, the total mass of the binary is taken as constant, because the effects of both mass loss and angular momentum loss are explicitly incorporated into Equations~\eqref{eq:orbit_averaged_semimajor_axis} and \eqref{eq:orbit_averaged_eccentricity} (see Section~3.3 in  \citetalias{2026A&A...706A..79P}).} we find
{\small
\begin{eqnarray}
  &\langle \frac{\dot{P}_{\rm orb}}{\dot{q}} \rangle = -\frac{3 P_{\rm orb}}{q(1+\beta q)} \frac{f_{a}(e,x)}{f_{\dot{M}_{\rm don}}(e,x)} \Biggl(1-\beta q-(1-\beta)\frac{(\gamma+\frac{1}{2})q}{1+q}\Biggr) \label{eq:dPdq},\\
  &\langle \frac{\dot{e}}{\dot{q}} \rangle = -\frac{2}{q(1+\beta q)} \frac{f_{e}(e,x)}{f_{\dot{M}_{\rm don}}(e,x)} \Biggl(1-\beta q-(1-\beta)\frac{(\gamma+\frac{1}{2})q}{1+q}\Biggr). \label{eq:dedq}
\end{eqnarray}
Equations~\eqref{eq:dPdq} and \eqref{eq:dedq} relate the secular evolution of $P_{\rm orb}$ and $e$ with that of $q$, and by dividing them, we find
\begin{equation}\label{eq:dedP}
    \langle \frac{\dot{e}}{\dot{P}_{\rm orb}} \rangle = \frac{2}{3P_{\rm orb}} \frac{f_{e}(e,x)}{f_{a}(e,x)}.
\end{equation}
}
Equations~\eqref{eq:dPdq} to \eqref{eq:dedP} define orbital evolutionary tracks (i.e., the evolution of the orbital elements during MT) and are independent of the orbit-averaged MT rate $\langle\dot{M}_{\rm don}\rangle$. Here, we assume for simplicity that the degree of RLOF is $x = 0.99$\footnote{During integration of Equations~\eqref{eq:orbit_averaged_semimajor_axis} and \eqref{eq:orbit_averaged_eccentricity} (i.e., as mass transfer progresses), the degree of RLOF parameter $x$ evolves self-consistently. Consequently, it affects the magnitude of the secular rates, but the evolution remains qualitatively unchanged (i.e., the signs do not change in Equations~\eqref{eq:dPdq} to \eqref{eq:dedP}).} and overplot the predicted orbital evolution from Equations~\eqref{eq:dPdq}, \eqref{eq:dedq}, and \eqref{eq:dedP} as teal tracks in Figures~\ref{fig:streamplot_dPdq}, \ref{fig:streamplot_dedq}, and \ref{fig:streamplot_dedP}, respectively. We illustrate Equations~\eqref{eq:dPdq} and \eqref{eq:dedq} for conservative MT ($\beta=1$), partially conservative MT ($\beta=0.5$) under isotropic reemission ($\gamma = q$), and nonconservative MT ($\beta=0$) assuming isotropic reemission ($\gamma = q$). Notably, Equation~\eqref{eq:dedP} is independent of both the accretion efficiency, $\beta$, and the AML efficiency, $\gamma$.

We start our discussion with Figures~\ref{fig:streamplot_dPdq} and \ref{fig:streamplot_dedq}. As $q$ decreases along the teal tracks during MT, the orbits initially shrink and tend to circularize until they reach the transitional mass ratio $q_{\rm trans}=1$ (conservative MT), $q_{\rm trans}=1.13$ (partially conservative MT under isotropic reemission), or $q_{\rm trans}=1.28$ (nonconservative MT under isotropic reemission), after which they widen and become more eccentric: $e$ and $P_{\rm orb}$ evolve in a correlated way. This behaviour is also evident across all models in Figures~\ref{fig:streamplot_dPdq} and \ref{fig:streamplot_dedq} (dashed, dotted and dash-dotted lines). Importantly, the observed wide and eccentric sdB+MS binaries all lie in the orbital-widening and eccentricity-pumping regime predicted by the GeMT model, consistent with systems that have completed MT after mass-ratio reversal.

The steepness of the teal tracks reflects the strength of the secular rates of change described by Equations~\eqref{eq:dPdq} and \eqref{eq:dedq}. For $q < q_{\rm trans}$ in all three MT cases, orbits evolve more strongly with decreasing $q$. This is shown by the $q$-dependence of Equations~\eqref{eq:dPdq} and \eqref{eq:dedq}. For instance, under conservative MT ($\beta = 1$) the equations simplify to
\begin{eqnarray}
  &\langle \frac{\dot{P}_{\rm orb}}{\dot{q}} \rangle = - 3 P_{\rm orb}\frac{(1-q)}{q(1+q)} \frac{f_{a}(e,x)}{f_{\dot{M}_{\rm don}}(e,x)} ,\\
  &\langle \frac{\dot{e}}{\dot{q}} \rangle = -2\frac{(1-q)}{q(1+q)} \frac{f_{e}(e,x)}{f_{\dot{M}_{\rm don}}(e,x)}, 
\end{eqnarray}
 and as $q \to 0$, both equations diverge to $+\infty$; this is expected since a less massive donor causes the semimajor axis to expand while eccentricity grows to conserve orbital angular momentum (\citetalias{2026A&A...706A..79P}). The behavior is the same for partially conservative and nonconservative MT, though $q_{\rm trans}$ is larger. Consequently, this $q$-dependence reflects angular momentum conservation and simultaneously explains the observed trends; systems with lower mass ratios tend to have higher eccentricities and longer orbital periods. 

Most importantly, Equations~\eqref{eq:dPdq} to \eqref{eq:dedP} also provide physical intuition for the origin and nature of the main and secondary branches. Equations~\eqref{eq:dPdq} and \eqref{eq:dedP} are explicitly dependent on $P_{\rm orb}$, thus evolutionary paths are shifted vertically (horizontally) on the $q - P_{\rm orb}$ ($P_{\rm orb} - e$) plane by the orbital period at RLOF onset, while exhibiting the same qualitative trends; as $q \to 0$ both $e$ and $P_{\rm orb}$ increase. As a result, secondary-branch progenitors must have initiated RLOF at shorter initial periods than main-branch ones. In contrast, Equation~\eqref{eq:dedq} is independent of $P_{\rm orb}$, implying that the post-MT eccentricity should be independent of the orbital period at which RLOF occurs. Interestingly, this is what we see in Figure~\ref{fig:streamplot_dedq}, where the distribution on the $q - e$ plane is unimodal, and the two branches collapse into a single group.

In summary, the GeMT model provides a theoretical interpretation of the orbital-parameter distributions of wide sdB+MS systems, and their observed correlations (1)-(3). Given the unimodal $q-e$ distribution (Figure~\ref{fig:streamplot_dedq}) and $P_{\rm orb}$-independence of Equation~\eqref{eq:dedq}, we propose that post-MT eccentricity is set primarily by MT physics itself (transferred mass, accretion efficiency, AML efficiency, etc.). This has profound implications: Not only does it provide an explanation of the unimodal distribution of wide sdB+MS on the $q - e$ plane, but it also indicates that eccentricity can be used as a probe to constrain MT physics.

\section{Discussion}\label{sec:four}

\subsection{Robustness of the orbital evolution trends}

\subsubsection{Tidal Interactions}

Throughout this work, we have focused on the dynamical affect of the MT alone, and neglected other processes such as tides. Here we discuss the role of tides before the onset of, and during, RLOF. Based on classical tidal theory \citep{1981A&A....99..126H,2002MNRAS.329..897H}, the orbits of Roche-lobe-filling stars have been expected to be circular. This prediction, however, is challenged both by more recent tidal formalisms and by observations. For instance, \citet{2021MNRAS.503.5569V,2025ApJ...984..137D} showed that binaries can initiate RLOF while still significantly eccentric, and interacting binaries with nonzero eccentricities have been known for decades \citep{1999AJ....117..587P,2005A&AT...24..151R}. In addition, with this new understanding of inefficient tides, the question is if tides are able to circularize any eccentricity that may be there during eccentric MT, but answering this question is not in the scope of our work. In summary, our results and insights are expected to remain valid as long as tidal interactions during eccentric MT have a secondary impact on the orbital evolution compared to the effects of eccentric MT itself.

\subsubsection{Stellar evolution}

A second effect that we have not taken into account is stellar evolution. Throughout this work we have assumed a constant donor radius $R_{\rm don}$ during MT. A changing $R_{\rm don}$ will alter only quantitatively the rates of change of the semimajor axis and eccentricity (see Equations~\eqref{eq:orbit_averaged_semimajor_axis} and \eqref{eq:orbit_averaged_eccentricity}) since $x \equiv R_{\rm L}^c/R_{\rm don}$, but the evolution remains qualitatively unchanged. Differences in the sdB progenitor evolution models and their microphysics (e.g., metallicity, overshooting) can affect the post-MT orbital parameters of our models. For a given sdB progenitor mass, the orbital period at which RLOF occurs is set by $R_{\rm don}$ at the TRGB. Moreover, the post-MT mass ratio (and thus the orbital period and eccentricity via Equations~\eqref{eq:dPdq} and \eqref{eq:dedq}) depends on $M_{\rm core}$ at the TRGB.  In our models, the sdB progenitor radii, $R_{\rm don}$, and helium core masses, $M_{\rm core}$, at the TRGB come from detailed MESA calculations by \citet{2023A&A...669A..45T}. Since both $R_{\rm don}$ and $M_{\rm core}$ can vary with metallicity and overshooting \citep{2016ApJ...823..102C}, different assumptions can affect the post-MT orbital parameters of the models. Despite such variations, the orbital evolution tracks and the overall paterns remain qualitatively unchanged (Equations~\eqref{eq:dPdq} to \eqref{eq:dedP}).

\subsection{Origin of the main and secondary branches}

An open question in the orbital-parameter distributions of wide sdB+MS binaries remains: What is the origin of the main and secondary branches (Figure~\ref{fig:streamplot_dPdq})? Based on our models (Section~\ref{sec:three}), the main branch can be linked to roughly $1$–$1.5~\mathrm{M}_{\odot}$ progenitors undergoing stable MT in systems with initial mass ratios of $q \lesssim 2.5$ depending on the accretion efficiency, $\beta$, and angular momentum loss, $\gamma$. The secondary branch must originate from systems with relatively heavier progenitors ($M \gtrsim 1.5~\mathrm{M}_{\odot}$) and/or relatively larger initial mass ratios (i.e., low-mass companions; Figure~\ref{fig:streamplot_dPdq}). This has three consequences: (1) more massive stars are increasingly rare according to the initial mass function \citep[IMF;][]{2001MNRAS.322..231K}, (2) increasingly large initial mass ratios, lead to unstable MT \citep[e.g.,][]{1987ApJ...318..794H,2010ApJ...717..724G,2015ApJ...812...40G,2020ApJ...899..132G}, and (3) the critical mass ratio for MT stability increases rapidly with progenitor mass \citep{2023A&A...669A..45T}. Consequently, such systems are likely to enter a common-envelope phase and either merge of end up at very short orbital periods if the envelope is ejected. We propose that the combination of these effects explains the scarcity of secondary-branch systems compared to those on the main branch. The fact that sdB+M-dwarf systems are only observed at short periods \citep[$0.05 \lesssim P_{\rm orb} \lesssim 1.6\, \mathrm{days}$][]{2019A&A...630A..80S,2022A&A...666A.182S} further supports this hypothesis.

\subsection{Eccentricity as a diagnostic of mass-transfer history}

\begin{figure}[!htbp]
    \centering
    \includegraphics[width=\linewidth]{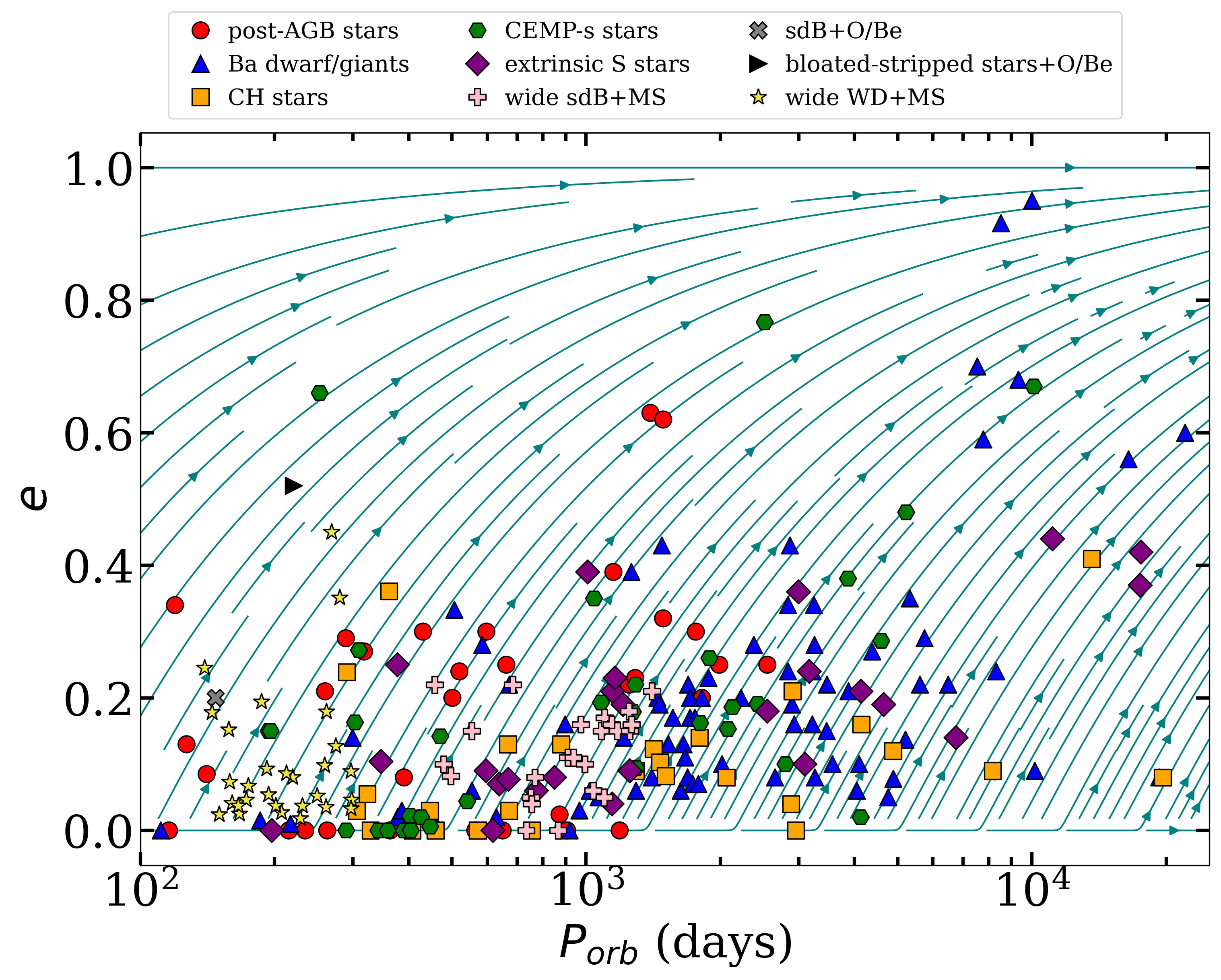}
    \caption{Similar to Figure~\ref{fig:streamplot_dedP}, but the data points now illustrate various wide post-interaction binaries. Red circles represent systems hosting post-AGB stars \citep{2018A&A...620A..85O}, blue triangles show Barium dwarfs and giants \citep{2019A&A...626A.127J,2019A&A...626A.128E,2020A&A...639A..24E}, orange squares show carbon-rich subgiants  \citep{2019A&A...626A.128E,2020A&A...639A..24E}, green pentagons illustrate carbon-enhanced metal-poor stars with s-process enhancement \citep{2016A&A...588A...3H,2016A&A...586A.158J,2016ApJ...826...85S}, purple diamonds show extrinsic S stars \citep{2000AJ....119.1375F,2019A&A...626A.127J,2020A&A...639A..24E}, pink plusses show sdB stars with main-sequence companions \citep{2012MNRAS.421.2798D,2012ApJ...758...58B,2012A&A...548A...6V,2013A&A...559A..54V,2017A&A...605A.109V,2019MNRAS.482.4592V,2021A&A...653A.120D,2022A&A...658A.122M}, and bright yellow stars represent white dwarfs with main-sequence companions \citep{2024MNRAS.52711719Y}. The gray cross symbol corresponds to 60 Cyg \citep{2024ApJ...962...70K}, the black triangle represents HIP 15429 \citep{2025A&A...701A...9M}.  
}
    \label{fig:period_eccentricity_plane}
\end{figure}

Besides wide sdB+MS systems, eccentric MT is relevant to a plethora of systems, ranging from those undergoing MT during the main-sequence to gravitational-wave progenitors. In Figure~\ref{fig:period_eccentricity_plane}, we present various wide post-interaction binaries with known orbital periods and eccentricities. Figure~\ref{fig:period_eccentricity_plane} is similar to Figure~\ref{fig:streamplot_dedP} and shows that the range of observed eccentricities increases with orbital period \citep[see also][their Figure 8]{2024MNRAS.529.3729S}; a pattern that is shared across nearly all system classes.

For orbits that initiate RLOF at nonzero eccentricities, the GeMT model predicts that widening orbits are accompanied by increasing eccentricities \citep[a qualitatively similar type of evolution is found by earlier eccentric-MT frameworks][]{2007ApJ...667.1170S,2009ApJ...702.1387S,2019ApJ...872..119H}, in agreement with observations. It is important to note that mechanisms such as CBD \citep{2015A&A...579A..49V,2020A&A...642A.234O} or triple interactions \citep{2016ComAC...3....6T,2016ARA&A..54..441N} may operate simultaneously with eccentric MT, requiring careful disentanglement of their effects on observed orbital-parameter distributions. Crucially, Equation~\eqref{eq:dedq} demonstrates that post-MT eccentricity is set primarily by MT physics itself (i.e. transferred mass, accretion efficiency $\beta$, AML efficiency, etc.). Figure~\ref{fig:period_eccentricity_plane} reveals that the observed post-interaction eccentricities vary among the different populations; for example, Barium stars span $0 \leq e < 1$, while wide sdB+MS systems generally have smaller eccentricities. We therefore highlight the potential of eccentricity measurements to constrain MT physics across post-interaction binary populations.

\section{Summary and conclusions}\label{sect:five}

In this letter, we used the GeMT framework 
\citep{2026A&A...706A..79P,2026A&A...706A.357P} to tackle the eccentricity problem and to explore the role of eccentric mass transfer in shaping the orbital parameters of post-mass-transfer binaries. To test the predictions of GeMT, we focused on sdBs with main-sequence companions in wide orbits, since their properties impose strong constraints on their pre-mass-transfer histories. Our main conclusions can be summarized as follows:
\begin{itemize}
    \item  When isolating the effects of MT via eccentric RLOF from other physical processes, GeMT explains simultaneously the observed orbital-parameter distributions and the correlations seen in wide sdB+MS binaries between $q_{\rm obs}$ and $P_{\rm orb}$, $q_{\rm obs}$ and $e$, and $P_{\rm orb}$ and $e$ (Section~\ref{sec:three}). To quantitatively constrain their orbital parameters, stellar evolution and tides should be included in future studies, but it is encouraging that eccentric mass transfer provides the first mechanism that qualitatively reproduces the observed distributions.
    \item By isolating eccentric mass transfer from other physical processes, we showed that the origin of the main and the secondary branches on the $q_{\rm obs} - P_{\rm orb}$  and $P_{\rm orb} - e$ planes in observed wide sdB+MS binaries can be traced back to systems initiating RLOF at sufficiently different orbital periods (Section~\ref{sec:three}). For instance, sdB progenitors with masses $M \sim 1-1.5~\mathrm{M}_{\odot}$ match the main branch, while heavier ones $M \gtrsim 1.5~\mathrm{M}_{\odot}$, which initiate RLOF at shorter periods, match the secondary branch (Figures~\ref{fig:streamplot_dPdq} and \ref{fig:streamplot_dedP}). 
    \item  We presented theoretical evidence (Equation~\eqref{eq:dedq}) that the eccentricity of post-mass-transfer systems is independent of the orbital period, but depends primarily on the details of mass transfer, such as the amount of transferred mass, the accretion efficiency, and the angular momentum loss. This finding also provides an explanation of the unimodal distribution of wide sdB+MS binaries on the $q_{\rm obs} - e$ plane (Figure~\ref{fig:streamplot_dedq}).
    \item Eccentricity is a common feature in multiple types of binaries, with components ranging from low- to high-mass stars as well as compact objects. Yet, eccentricity has often been discarded in binary-evolution studies. Our results instead show that post-mass-transfer eccentricities are directly shaped by the underlying mass-transfer physics and thus encode information about the formation pathways of these systems. Since these binaries are progenitors of some of the most energetic astrophysical phenomena, including gravitational-wave sources and SNIa-like transients, eccentricity emerges as a key observable diagnostic for constraining binary evolution and, in turn, for understanding the origin of these events.
\end{itemize}

\section{Data availability}

The data necessary to reproduce the models presented in Figures~\ref{fig:streamplot_dPdq} to \ref{fig:streamplot_dedP} in this letter is available on Zenodo. The GeMT code will be shared upon reasonable request to the authors.

\begin{acknowledgements}
AP would like to thank Philipp Podsiadlowski for suggesting the wide sdB binaries as a post-mass-transfer population to test mass-transfer phyiscs. AP would like to thank Onno Pols for the useful discussions. The authors would like to thank Karel Temmink for providing stellar parameters from his detailed MESA models \citep{2023A&A...669A..45T}. AP \& ST acknowledges support from the Netherlands Research Council NWO (VIDI 203.061 grant). EL acknowledges support through a start-up grant from the Internal Funds KU Leuven (STG/24/073), through a Veni grant (VI.Veni.232.205) from the Netherlands Organization for Scientific Research (NWO), and the Research Foundation – Flanders (FWO) under the Odysseus Program, Type II (G0AT525N). VS thanks Hongwei Ge and the binary group at Yunnan observatories in Kunming, China, for the invitation to the conference "Binary Stars in a new Era", which made this collaboration possible. This work used the following software packages: Matplotlib \cite{Hunter:2007}, NumPy \cite{harris2020array}, SciPy \citep{2020SciPy-NMeth} and SymPy \citep{10.7717/peerj-cs.103}.
\end{acknowledgements}

\bibliography{references}
\bibliographystyle{aasjournalv7}

\begin{appendix}

\section{The initial eccentricity}\label{app:sensitivity}

For the models presented in Section~\ref{subsec:comparison}, we adopted a seed eccentricity of $e = 10^{-3}$ at the onset of RLOF, since such seed eccentricities are expected for Roche-lobe-filling giants \citep{1992RSPTA.341...39P,2024MNRAS.534..455C}. Because, the endpoints of the evolution of mass-transferring binaries depend on the initial eccentricity via Equation~\eqref{eq:dedP}, we perform a sensitivity test on the GeMT-model predictions. We construct a grid of initial conditions as follows. For each sdB progenitor mass ($M_{\rm don} = 1\,\mathrm{M}_{\odot}$, $1.5\,\mathrm{M}_{\odot}$, and $2\,\mathrm{M}_{\odot}$), we sample 10 initial accretor masses ($M_{\rm acc}$) uniformly between $[M_{\rm don}/q_{\rm crit},\, M_{\rm don}-0.1]\,\mathrm{M}_{\odot}$, where $q_{\rm crit} = 3$. The lower limit approximates the minimum accretor mass required for stable mass transfer, while the upper limit ensures that the donor is initially the more massive star in the system. For each of these 30 systems, we assume three initial eccentricities: $e = 10^{-3}$, $10^{-4}$, and $10^{-5}$. Following the same procedure as in Section~\ref{subsec:comparison}, we track the orbital evolution under (a) conservative MT ($\beta = 1$), (b) partially conservative MT ($\beta = 0.5$), and (c) non-conservative MT ($\beta = 0$), with the latter two assuming isotropic reemission ($\gamma = q$). In summary, for each sdB progenitor mass we obtain a grid of 90 post-MT systems.

In each grid, 37 of the 90 systems terminate during MT irrespective of initial eccentricity, indicating that this outcome is governed by the mass ratio rather than by the initial eccentricity. Of these unsuccessful systems, most (34/37) terminate because the orbit becomes effectively parabolic ($e > 0.99$), while the remaining (3/37) terminate due to merger. Among the 53 systems that could be compared directly, the largest differences reach $3.5\times10^{4}$ days in orbital period and 0.056 in eccentricity, but these extremes are confined to very low mass ratios ($q < 0.3$) and/or very high eccentricities ($e \gtrsim 0.9$). Importantly, none of the observed wide sdB+MS binaries has $q \lesssim 0.3$ or $e \gtrsim 0.95$, so this highly sensitive regime is not relevant for the systems we compare to.

Overall, the impact of initial eccentricity on the final orbital parameters is modest. For these 53 systems, as the initial eccentricity is lowered from $10^{-3}$ to $10^{-4}$ and $10^{-5}$, the median absolute relative change in the post-MT orbital period is 0.1\%–0.2\% (1.6–1.8 days), while the median absolute relative change in the post-MT eccentricity is 5.7\%–7.6\% (0.004–0.005). Thus, for a typical system the predicted post-MT period and eccentricity differ by only $\sim$1.6–1.8 days and $\sim$0.004–0.005, respectively. We conclude that, within this range, the choice of initial eccentricity has a negligible effect on the bulk population of final periods and eccentricities, and that the orbital evolution and its trends remain qualitatively unchanged for models relevant to observed wide sdB+MS binaries.

\section{Supplementary tables}\label{app:tables}

\begin{table}[!htbp]
\caption{Orbital parameters of the wide sdB+MS binary sample.}\label{tab:observables}
  \centering
  \begin{tabular}{cccccc}
  \hline
  Object& Branch & $P_{\rm orb}$ & $e$ & $q_{\rm obs}$ & Reference \\
  & &(days) & & & \\
  \hline \hline
    EC22536-5304 & secondary & 457 $\pm$ 1.5 & 0.22 $\pm$ 0.08 & 0.69 $\pm$ 0.05 & (1) \\
    PG 1514+034 & secondary & 479 $\pm$ 2& 0.1 $\pm$ 0.02 & 0.58 $\pm$ 0.03 & (2)\\
    BD$-$11\degr162 &  secondary & 497.1 $\pm$ 0.2 & 0.082 $\pm$ 0.002 & 0.48 $\pm$ 0.01 & (3)\\
    GALEX J022836.7$-$362543 & secondary & 554 $\pm$ 1 & 0.15 $\pm$ 0.02 & 0.5 $\pm$ 0.08 & (2)\\
    PB 6355 & secondary & 684 $\pm$ 31 & 0.22 $\pm$ 0.06 & 0.32 $\pm$ 0.02 & (2)\\
    PG 1018$-$047& main & 752 $\pm$ 2 & 0.05 $\pm$ 0.01 & 0.7 $\pm$ 0.02 & (4)\\
    PG 1104+243 & main &755 $\pm$ 5 & 0.04 $\pm$ 0.02 & 0.71 $\pm$ 0.02 & (5)\\
    MCT 0146-2651 & main & 768 $\pm$ 11 & 0.08 $\pm$ 0.06 & 0.66 $\pm$ 0.03 & (2)\\
    GALEX J053939.1$-$283329 & main & 865 $\pm$ 6 &  0 & 0.74 $\pm$ 0.09 & (2)\\
    PG 1149+653 & main & 909 $\pm$ 2& 0.11 $\pm$ 0.02 & 0.64 $\pm$ 0.07 & (6)\\
    Feige 87 & main & 936 $\pm$ 2& 0.11 $\pm$ 0.01 & 0.55 $\pm$ 0.01 & (7)\\
    BD+34\degr1543 & main & 972 $\pm$ 2& 0.16 $\pm$ 0.01 & 0.57 $\pm$ 0.01 & (7)\\
    FAUST 321 & main & 993 $\pm$ 15& 0.1 $\pm$ 0.03 & 0.45 $\pm$ 0.01 & (2)\\
    EC 03143$-$5945 & main & 1037 $\pm$ 3 & 0.06 $\pm$ 0.02 & 0.41 $\pm$ 0.02 & (2)\\
    JL 277 & main & 1082 $\pm$ 9& 0.15 $\pm$ 0.04 & 0.42 $\pm$ 0.02 & (2)\\
    TYC 2084$-$448$-$1 & main & 1098 $\pm$ 5& 0.05 $\pm$ 0.03 & 0.51 $\pm$ 0.02 & (8)\\
    EC 11031$-$1348 & main & 1099 $\pm$ 6& 0.17 $\pm$ 0.03 & 0.36 $\pm$ 0.02 & (8)\\
    Feige 80 & main & 1140.4 $\pm$ 5& 0.16 $\pm$ 0.02 & 0.42 $\pm$ 0.02 & (3)\\
    GALEX J162842.0+111838 & main & 1176 $\pm$ 30& 0.15 $\pm$ 0.05 & 0.42 $\pm$ 0.04 & (2)\\
    GALEX J033216.7$-$023302 & main & 1247 $\pm$ 30& 0.18 $\pm$ 0.05 & 0.36 $\pm$ 0.07 & (2)\\
    BD+29\degr3070 & main & 1254 $\pm$ 5& 0.15 $\pm$ 0.01 & 0.37 $\pm$ 0.02 & (8)\\
    BD$-$07\degr5977 & main & 1262 $\pm$ 1& 0.16 $\pm$ 0.01 & 0.44 $\pm$ 0.1 & (8)\\
    TYC 3871$-$835$-$1 & main & 1263 $\pm$ 5& 0.16 $\pm$ 0.02 & 0.54 $\pm$ 0.02 & (8)\\
    PG 2148+095 & main & 1404 $\pm$ 92& 0.21 $\pm$ 0.06 & 0.34 $\pm$ 0.06 & (2)\\
  \hline
  \end{tabular}
  \tablecomments{From left to right, object's identification, orbital period, eccentricity, mass ratio defined as the mass of the sdB over the mass of the main-sequence companion, and the reference to the source article.\\
  \textbf{References.}
  (1) \citet{2021A&A...653A.120D};
  (2) \citet{2019MNRAS.482.4592V};
  (3) \citet{2022A&A...658A.122M};
  (4) \citet{2018MNRAS.474..433D};
  (5) \citet{2012A&A...548A...6V};
  (6) \citet{2012ApJ...758...58B};
  (7) \citet{2013A&A...559A..54V};
  (8) \citet{2017A&A...605A.109V}.}
\end{table}

\begin{table}[!htbp]
\caption{Initial conditions at the onset of RLOF.}\label{tab:integration_parameters}
  \centering
  \begin{tabular}{ccccccccc}
  \hline
  Model & $M_{\rm don}$ & $M_{\rm acc}$ & $P_{\rm orb}$ & $e$ & $\mathrm{R}_{\rm don}$ (TRGB) & x & $\beta$ & $\gamma$\\
  & (M$_{\odot}$) & (M$_{\odot}$) & (days) & &(R$_{\odot}$)& &\\
  \hline \hline
  1.0+0.7+C & 1.0 & 0.7 & 803.7 &  0.001 & 180 & 0.99 & 1.0 & -\\
  1.0+0.7+PC & 1.0 & 0.6 & 803.7 &  0.001 & 180 & 0.99 & 0.5 & q\\
  1.0+0.7+NC & 1.0 & 0.5 & 803.7 &  0.001 & 180 & 0.99 & 0.0 & q\\
  \hline
  1.5+0.8+C & 1.0 & 0.8 & 505.47 &  0.001 & 155 & 0.99 & 1.0 & -\\
  1.5+0.7+C & 1.0 & 0.7 & 498.04 &  0.001 & 155 & 0.99 & 1.0 & -\\
  1.5+0.6+C & 1.0 & 0.6 & 486.05 &  0.001 & 155 & 0.99 & 1.0 & -\\
  1.5+0.8+PC & 1.5 & 0.8 & 505.47 &  0.001 & 155 &  0.99 & 0.5 & q\\
  1.5+0.7+PC & 1.5 & 0.7 & 498.04 &  0.001 & 155 &  0.99 & 0.5 & q\\
  1.5+0.6+PC & 1.5 & 0.6 & 486.05 &  0.001 & 155 &  0.99 & 0.5 & q\\
  1.5+0.8+NC & 1.5 & 0.8 & 505.47 &  0.001 & 155 &  0.99 & 0.0 & q\\
  1.5+0.7+NC & 1.5 & 0.7 & 498.04 &  0.001 & 155 &  0.99 & 0.0 & q\\
  1.5+0.6+NC & 1.5 & 0.6 & 486.05 &  0.001 & 155 &  0.99 & 0.0 & q\\
  \hline
  2.0+0.7+C & 1.0 & 0.7 & 153.02 &  0.001 & 80 & 0.99 & 1.0 & -\\
  2.0+0.7+PC & 1.0 & 0.6 & 153.02 &  0.001 & 80 & 0.99 & 0.5 & q\\
  2.0+0.7+NC & 1.0 & 0.5 & 153.02 &  0.001 & 80 & 0.99 & 0.0 & q\\
  \hline
  \end{tabular}
  \tablecomments{Models are labeled by the donor mass $M_{\rm don}$, the accretor mass $M_{\rm acc}$, and and whether MT is conservative (C), partially conservative (PC) or nonconservative (NC). $\mathrm{R}_{\rm don}$ corresponds to the radius of the donor star near the tip of the red giant branch (TRGB).} 
\end{table}

\end{appendix}
\end{document}